\documentclass[twoside,english,aps,prl,twocolumns,reprint,showpacs,preprintnumbers,superscriptaddress,nofootinbib,floatfix,10pt]{revtex4-1}
\usepackage[T1]{fontenc}
\usepackage[latin9]{inputenc}
\usepackage{color}
\usepackage{babel}
\usepackage{bm}
\usepackage{amsmath}
\usepackage{amsthm}
\usepackage{amssymb}
\usepackage{graphicx}
\usepackage[unicode=true,
 bookmarks=true,bookmarksnumbered=false,bookmarksopen=false,
 breaklinks=false,pdfborder={0 0 0},pdfborderstyle={},backref=false,colorlinks=true]
 {hyperref}
\hypersetup{pdftitle={SU4Interaction},
 pdfauthor={BNLU},
 unicode=true, bookmarks=true,bookmarksnumbered=false,bookmarksopen=false, breaklinks=false,pdfborder={0 0 1},backref=false,colorlinks=true, allcolors=blue}

\makeatletter
\usepackage{amsfonts,amssymb,latexsym,amsmath}
\usepackage{graphicx}
\usepackage{times}
\usepackage{slashed}
\allowdisplaybreaks[4]

\makeatother

\begin{document}
\title{Renormalization of many-body effective field theory}
\author{Bing-Nan Lu}
\email{bnlv@gscaep.ac.cn}

\affiliation{Graduate School of China Academy of Engineering Physics, Beijing 100193,
China}
\author{Bao-Ge Deng}
\affiliation{Institute of Theoretical Physics, Chinese Academy of Sciences, Beijing
100190, China}
\affiliation{School of Physical Sciences, University of Chinese Academy of Sciences,
Beijing 100049, China}
\begin{abstract}
The renormalization of the effective field theories (EFTs) in many-body
systems is the most pressing and challenging problem in modern nuclear
\textit{ab initio} calculation. For general non-relativistic EFTs,
we prove that the renormalization group (RG) invariance can be achieved
if and only if all \textit{single-particle} momenta are regulated
with a universal cutoff $\Lambda$. For a numerical demonstration,
we construct a series of N$^{2}$LO chiral forces with $\Lambda$
varying from $250$ MeV to $400$ MeV. With all low energy constants
fixed in two- and three-nucleon systems, we reproduce the experimental
binding energies of $^{4}$He and $^{16}$O nearly independently of
$\Lambda$. In contrast, all recent nuclear EFT constructions regulate
the \textit{relative} momenta for Galilean invariance, thus inherently
break the RG invariance. This explains the unpleasantly strong cutoff-dependences
observed in recent \textit{ab initio} calculations. Our method can
also be used to build RG-invariant EFTs with non-perturbative interactions.
\end{abstract}
\maketitle

\paragraph{Introduction}

The chiral effective field theory ($\chi$EFT), usually conceived
as a low-energy alternative to the QCD, has become the standard tool
for nuclear \textit{ab initio} calculations\cite{Epelbaum2009review,Machleidt2011review,Epelbaum2020review}.
However, recently this paradigm faces a great crisis in nuclear many-body
calculations. As in the cases of the conventional nuclear forces,
it is possible to build different $\chi$EFT constructions with the
same quality of the nucleon-nucleon scattering phase shifts and few-body
observables\cite{Entem2003PRC,Epelbaum2005NPA,Epelbaum2015PRL,Hebeler2015PRC,Entem2017PRC,Reinert2018EPJA}.
When applied to heavier nuclei or nuclear matter, however, their predictions
are not unique and rather model-dependent.  More specifically,
we take the phase-equivalent $\chi$EFTs generated with the momentum
similarity renormalization group (SRG) method\cite{Bogner2010review,Bogner2007PRC,Anderson2008PRC,Gubankova1998,Gubankova2000}
as an example. The momentum SRG method transforms the interactions
via a series of unitary transformations parametrized by a SRG flow
parameter $\lambda$\cite{Bogner2007PRC} or running momentum cutoff
$\Lambda$\cite{Anderson2008PRC,Gubankova1998,Gubankova2000}. With
the SRG all evolved two-nucleon forces (2NFs) give exactly the same
low-momentum phase shifts. However, when supplemented with the three-nucleon
forces (3NFs) fixed in three-nucleon systems\cite{Hebeler2011PRC,Hebeler2012PRC,Hebeler2021review},
these SRG-evolved interactions make significantly different predictions
for heavier nuclei\cite{Simonis2017PRC} or nuclear matter\cite{Hebeler2011PRC,Drischler2019PRL}.
For example, the binding energy of $^{40}$Ca can vary by about $20$\%
when $\Lambda(\lambda)$ is varied by $10\%$\cite{Simonis2017PRC}.
Such a strong dependence on $\Lambda$($\lambda$) indicates a severe
violation of the renormalization group (RG) invariance. So far the
underlying mechanism has not been understood. As a pragmatic solution,
some $\chi$EFT constructions are optimized for many-body calculations
by adjusting $\Lambda(\lambda)$ as a hidden parameter\cite{Hebeler2011PRC,Hebeler2021review,Drischler2019PRL,Hoppe2019PRC,Huther2020PLB,Kim2018NTSE,Papak2021JPG}.
Nevertheless, such fine-tunings can not be justified within the standard
EFT paradigm and would become impossible for a fully RG-invariant
theory.

Here we present an explanation for the observed strong RG-invariance
breaking effects. We note that there is a mostly unnoticed inconsistency
in these SRG-evolved forces, \textit{i.e.}, the two- and three-body
forces are evolved in the reduced Hilbert spaces of the relative coordinates,
then transferred to the many-body spaces by assuming the Galilean
invariance. We already know that these transferred forces are not
equivalent, otherwise we would have obtained the same many-body observables.
 In this sense the RG invariance and the Galilean invariance are
mutually incompatible. One solution is to relax the strict Galilean
invariance and build an EFT that consistently handles the two-, three-
and many-body Hilbert spaces on an equal footing. Then the RG invariance
in two- and three-body spaces would imply the RG invariance in many-body
spaces, and \textit{vice versa}. In this Letter we present such an
EFT construction, then prove its RG invariance by explicitly building
the unitary transformations between the theories defined at different
scales. We also give a numerical demonstration using realistic chiral
forces.

\paragraph{Decomposition of the Fock space}

To illustrate the idea, we start with an EFT of spinless bosons. The
Hamiltonian is
\begin{equation}
H=\int d\tau:-\frac{\Phi^{\dagger}\nabla^{2}\Phi}{2m}+\frac{C_{2}}{2}(\Phi_{\Lambda}^{\dagger}\Phi_{\Lambda})^{2}+\frac{C_{3}}{6}(\Phi_{\Lambda}^{\dagger}\Phi_{\Lambda})^{3}+...:,\label{eq:Hamiltonian}
\end{equation}
where $\Phi$ ($\Phi^{\dagger})$ is the annilation (creation) operator,
$m$, $C_{2,3}$ are particle mass and coupling constants, respectively.
The colons denote the normal ordering. The elipsis stand for higher-body
forces and interactions containing spatial derivatives. The subscript
$\Lambda$ means that the interactions only act on particles with
momentum lower than $\Lambda$,
\begin{equation}
\Phi_{\Lambda}^{\dagger}(\bm{r})=\int\frac{d^{3}p}{(2\pi)^{3}}\theta(\Lambda-|\bm{p}|)\exp(i\bm{p}\cdot\bm{r})a_{\bm{p}}^{\dagger},\label{eq:op_cutoff}
\end{equation}
where $a_{\bm{p}}^{\dagger}$ is the creation operator in momentum
representation, $\theta$ is the step function serving as a momentum
regulator. $\Phi_{\Lambda}^{\dagger}$ create a particle with a finite
extension $\sim\Lambda^{-1}$. Note that the cutoff scheme in Eq.(\ref{eq:op_cutoff})
is imposed on single-particle momenta instead of relative momenta,
thus the Hamiltonain Eq.(\ref{eq:Hamiltonian}) explicitly breaks
the Galilean invariance.

We now turn to the Fock space spanned by momentum eigenstates $a_{\bm{p}_{1}}^{\dagger}a_{\bm{p}_{2}}^{\dagger}...a_{\bm{p}_{N}}^{\dagger}|0\rangle$
with $|0\rangle$ the vacuum. $N\in[0,\infty)$ is the particle number.
In this basis the Fock space can be decomposed into a direct sum of
two subspaces. We define space-$0$ as spanned by momentum basis with
all $|\bm{p}_{i}|\leq\Lambda$, while space-$1$ is spanned by basis
with at least one $|\bm{p}_{i}|>\Lambda$. For any state $|\phi\rangle$
in space-$0$, $H|\phi\rangle$ has no particle excited to levels
above $\Lambda$ and also belongs to space-$0$. Thus $H$ does not
couple the two subspaces and we can simply drop space-$1$ without
affecting all low-momentum physics in space-$0$.

\begin{figure}
\begin{centering}
\includegraphics[width=0.9\columnwidth]{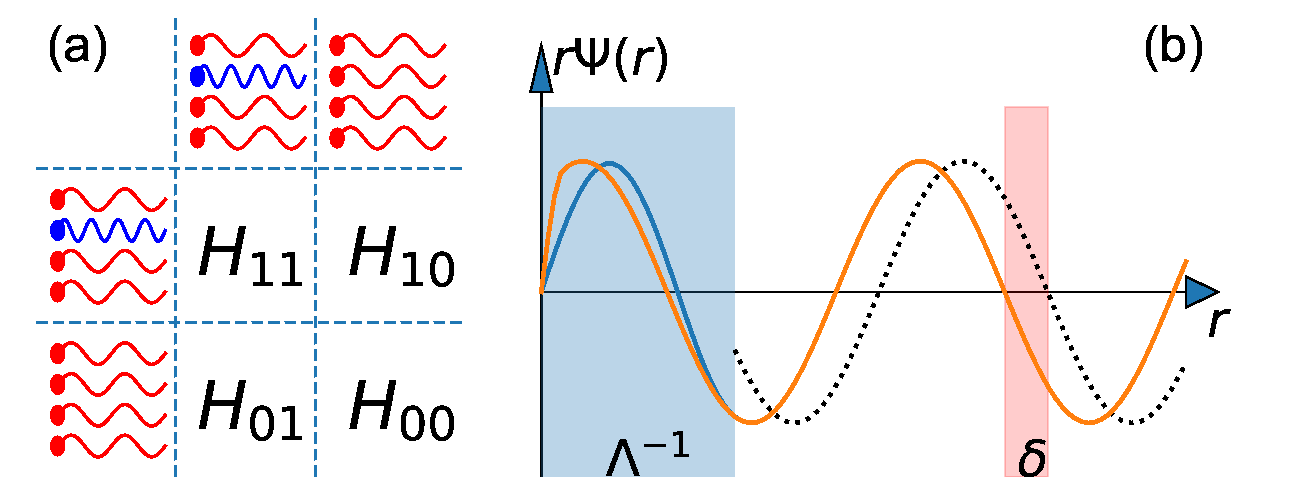}
\par\end{centering}
\caption{\label{fig:schematic}(a) Schematic plot of the partitioning of the
Hamiltonian in the Fock space. Red (blue) points denote particles
with momentum $|\bm{p}|\protect\leq\Lambda^{\prime}$ ($\Lambda^{\prime}<|\bm{p}|\protect\leq\Lambda$).
(b) Schematic plot of radial wave functions for two-particle scattering.
Red (Blue) lines denote the results calculated with $H$ ($H^{\prime})$.
In asymptotic region $r\gg\Lambda^{-1}$ the two wave functions coincide
and their difference from the free wave function (dotted line) gives
the phase shift $\delta$.}
\end{figure}

We next set a new cutoff $\Lambda^{\prime}$ slightly lower than $\Lambda$.
Again we can decompose space-$0$ into a direct sum of a space-$0^{\prime}$
spanned by states with all $|\bm{p}_{i}|\leq\Lambda^{\prime}$ and
a space-$1^{\prime}$ spanned by states with at least one momentum
$\bm{p}_{i}$ satisfying $\Lambda^{\prime}<|\bm{p}_{i}|\leq\Lambda$.
This corresponds to a partition of $H$ as schematized in Fig.\ref{fig:schematic}(a).
To see the exact form of the partitioned Hamiltonians, we split the
field operator as $\Phi_{\Lambda}^{\dagger}=\Phi_{\Lambda^{\prime}}^{\dagger}+\delta\Phi^{\dagger}$,
where $\Phi_{\Lambda^{\prime}}^{\dagger}$ is defined by Eq.(\ref{eq:op_cutoff})
with $\Lambda$ substituted by $\Lambda^{\prime}$. The operator $\Phi_{\Lambda^{\prime}}^{\dagger}$
creates the soft modes and $\delta\Phi^{\dagger}$ creates the hard
modes we shall remove. Then we can rewrite $H$ into an expansion
in $\Phi_{\Lambda^{\prime}}^{\dagger}$, $\delta\Phi^{\dagger}$ and
their conjugates. The first few terms are 
\begin{eqnarray}
H_{0} & = & \int d\tau:-\frac{\Phi^{\dagger}\nabla^{2}\Phi}{2m}+\frac{C_{2}}{2}(\Phi_{\Lambda^{\prime}}^{\dagger}\Phi_{\Lambda^{\prime}})^{2}+\frac{C_{3}}{6}(\Phi_{\Lambda^{\prime}}^{\dagger}\Phi_{\Lambda^{\prime}})^{3}\nonumber \\
 &  & +2C_{2}(\Phi_{\Lambda^{\prime}}^{\dagger}\Phi_{\Lambda^{\prime}})(\delta\Phi^{\dagger}\delta\Phi)+C_{2}(\Phi_{\Lambda^{\prime}}^{\dagger}\delta\Phi)(\delta\Phi^{\dagger}\delta\Phi)\nonumber \\
 &  & +\frac{3C_{3}}{2}(\Phi_{\Lambda^{\prime}}^{\dagger}\Phi_{\Lambda^{\prime}})^{2}\delta\Phi^{\dagger}\delta\Phi+\frac{C_{3}}{2}(\Phi_{\Lambda^{\prime}}^{\dagger}\delta\Phi)^{2}\delta\Phi^{\dagger}\delta\Phi+...:,\nonumber \\
H_{1} & = & \int d\tau:C_{2}(\Phi_{\Lambda^{\prime}}^{\dagger}\Phi_{\Lambda^{\prime}})(\Phi_{\Lambda^{\prime}}^{\dagger}\delta\Phi)+\frac{C_{2}}{2}(\delta\Phi^{\dagger}\Phi_{\Lambda^{\prime}})^{2}\nonumber \\
 &  & +\frac{C_{3}}{2}(\Phi_{\Lambda^{\prime}}^{\dagger}\Phi_{\Lambda^{\prime}})^{2}\Phi_{\Lambda^{\prime}}^{\dagger}\delta\Phi+\frac{C_{3}}{6}(\delta\Phi^{\dagger}\Phi_{\Lambda^{\prime}})^{3}+...:,\label{eq:H_decomposition}
\end{eqnarray}
where $H_{0}$ and $H_{1}$ are both Hermite and $H=H_{0}+H_{1}$.
$H_{0}$ is composed of all diagonal terms defined as containing no
$\delta\Phi^{\dagger}$ and $\delta\Phi$ or containing both $\delta\Phi^{\dagger}$
and $\delta\Phi$, whereas $H_{1}$ consists of all residual terms
containing $\delta\Phi$ or $\delta\Phi^{\dagger}$ but not both of
them. For any state $|\phi\rangle$ in space-$0^{\prime}$, we see
that $H_{0}|\phi\rangle$ only contains low-momentum particles with
$|\bm{p}|\leq\Lambda^{\prime}$ and belongs to space-$0^{\prime}$,
and $H_{1}|\phi\rangle$ has one or more particles excited to levels
above $\Lambda^{\prime}$ and belongs to space-$1^{\prime}$. Thus
$H_{0}$ is block-diagonal and vanishes in the off-diagonal regions
$H_{01}$ and $H_{10}$ in Fig.\ref{fig:schematic}(a), while $H_{1}$
vanishes in the region $H_{00}$. 

It is straightforward to verify that the commutator of any two terms
in Eq.(\ref{eq:H_decomposition}) can be written as a linear combination
of several such terms. The key point is to rewrite the commutator
into a normal-ordered series using Wick's theorem, then the terms
without contractions cancel out, every remaining term contains at
least one short-range $\delta$-like function from the contractions
and can be expressed using the contact terms. In mathematical terminology,
we say that the terms in Eq.(\ref{eq:H_decomposition}) span an infinite
dimensional Lie algebra.

\paragraph{Decoupling the soft and hard modes}

Next we search for a unitary transformation that decouples space-$0^{\prime}$
and space-$1^{\prime}$ by eliminating the off-diagonal blocks of
the Hamiltonian. For this purpose it is convenient to use the SRG
flow equation\cite{Anderson2008PRC,Gubankova1998,Gubankova2000}.
This method involves a time-dependent unitary transformation $U(t)$
with $t\in[0,\infty)$. The transformed Hamiltonian $H(t)=U(t)^{-1}HU(t)$
follows a flow equation
\begin{equation}
i\partial_{t}H(t)=[\eta(t),H(t)],\label{eq:SRGflowequation}
\end{equation}
where the operator $\eta(t)=i[H_{0}(t),H_{1}(t)]$ is a Hermite generator
satisfying $i\partial_{t}U(t)=\eta(t)U(t)$. By computing the commutators
explicitly we see that the flow defined by Eq.(\ref{eq:SRGflowequation})
amounts to a continuous evolution of the coefficients in Eq.(\ref{eq:H_decomposition}).
As long as $H_{1}(t)$ is not too large, we expect that the evolution
suppresses the off-diagonal matrix elements in $H_{1}(t)$\cite{SupMat}.
For $t\rightarrow\infty$ all terms in $H_{1}(t)$ vanish, the flow
converges to a fixed point $H_{1}=\eta=0$ and $U(t)$ has a well-defined
limit $U(\infty)$. The resulting Hamiltonian $H(\infty)=U(\infty)HU(\infty)^{-1}$
only contains diagonal terms. We can further apply a projector $P_{0^{\prime}}$
to space-$0^{\prime}$ to remove all terms containing both $\delta\Phi$
and $\delta\Phi^{\dagger}$. The final result $H^{\prime}=P_{0^{\prime}}H(\infty)P_{0^{\prime}}$
has the same form as Eq.(\ref{eq:Hamiltonian}) with all $\Lambda$
substituted by $\Lambda^{\prime}$ and all coefficients $m$, $C_{2}$,
$C_{3}$, etc. updated to new values $m^{\prime}$, $C_{2}^{\prime}$,
$C_{3}^{\prime}$, etc. Because $U(\infty)$ is unitary, the Hamiltonians
$H^{\prime}$ and $H$ have exactly the same low-lying spectra. 

Several important conclusions can be immediately drawn by inspecting
the flow Eq.(\ref{eq:SRGflowequation}). First, as the generator $\eta(t)$
is a spatial integral of local operators consisting of equal numbers
of creation and annilation opeartors, both $\eta(t)$ and the induced
transformation $U(\infty)$ conserves the particle number $N$ and
total momentum $\bm{P}$. Consequently, the single particle state
$|\bm{p}\rangle=a_{\bm{p}}^{\dagger}|0\rangle$ with $|\bm{p}|<\Lambda^{\prime}$
is a common eigenvector of $H$ and $H^{\prime}$ with the same eigenvalue
$E=p^{2}/2m$. It follows immediately that $m^{\prime}=m$ and other
one-body terms like $\Phi^{\dagger}\nabla^{4}\Phi$ are strictly forbidden.
 Second, as $\eta(t)$ does not contain any long-range interaction,
the transformation $U(\infty)$ is localized that $U(\infty)(|\phi_{1}\rangle\otimes|\phi_{2}\rangle)=U(\infty)|\phi_{1}\rangle\otimes U(\infty)|\phi_{2}\rangle$
where $|\phi_{1,2}\rangle$ are single- or many-particle wave packets
separated by a distance much larger than $\Lambda^{-1}$ and the symbol
$\otimes$ means the tensor product. For a typical reaction, the incoming
(outgoing) state long before (after) the collision consists of well
seperated clusters, each of which corresponds to a bound state of
the Hamiltonian. Applying the transformation $U(\infty)$ to both
the Hamiltonian and the incoming (outgoing) state, we transform $H$
to $H^{\prime}$ and the cluster wave functions from eigenvectors
of $H$ to that of $H^{\prime}$ with the same binding energies, meanwhile
the unitarity of $U(\infty)$ ensures the invariance of the $S$-matrix
element. In Fig.\ref{fig:schematic}(b) we schematically compare the
radial wave functions for a two-particle scattering process. The localized
transformation $U(\infty)$ modifies the short-range wave function
at $r\lesssim\Lambda^{-1}$ but preserves the asymptotic behaviour
and the phase shift. For more complicated inelastic scattering and
reactions, no such simple picture is available, but our arguments
based on spatial locality is still tenable.

In summary, the existence of the localized unitary transformation
$U(\infty)$ hinges on the presence of the direct-sum decomposition
of the Fock space, which in turn relies on the usage of a universal
single-particle regulator. In contrast, all recent chiral EFT constructions
employ Galilean invariant regulators acting on relative momenta or
even use different cutoffs for two- and three-body contact terms.
Such regulators can not unambiguously and consistently separate the
low- and high-momentum many-body Hilbert spaces. In these cases the
RG invariance of the energy spectrum and $S$-matrix elements is not
guaranteed by any known mechanism, and their cutoff-dependences are
mostly uncontrollable.

The above procedure of lowering the cutoff can be repeated many times
to generate the RG flows, which can then be analyzed using the whole
toolbox of the conventional Wilsonian RG theory. We note that in most
applications we do not need to explicitly compute hundreds of commutators
in Eq.(\ref{eq:SRGflowequation}). It suffices to write down the most
general Hamiltonian with the correct symmetries and regulators, then
determine the LECs by matching to low-momentum observables.

\paragraph{RG invariance of the chiral EFT}

The above discussions can be easily generalized to spin-1/2 Fermions
like the nucleons. For a numerical demonstration, we take a simplified
N$^{2}$LO chiral Hamiltonian $H_{{\rm N2LO}}=K+V_{{\rm 2N}}+V_{{\rm 3N}}+V_{{\rm C}}+V_{{\rm OPEP}}$,
where $K$ is the kinetic energy term, $V_{{\rm 2N}}$ and $V_{{\rm 3N}}$
are two- and three-body contact terms, $V_{{\rm C}}$ and $V_{{\rm OPEP}}$
are Coulomb and one-pion-exchange potentials, respectively\cite{Lu2022PRL-1}.
In momentum space the two-nucleon force (2NF) is written as
\begin{eqnarray*}
V_{{\rm 2N}} & = & \left[B_{1}+B_{2}(\bm{\sigma}_{1}\cdot\bm{\sigma}_{2})+C_{1}q^{2}+C_{2}q^{2}(\bm{\tau}_{1}\cdot\bm{\tau}_{2})\right.\\
 &  & +C_{3}q^{2}(\bm{\sigma}_{1}\cdot\bm{\sigma}_{2})+C_{4}q^{2}(\bm{\sigma}_{1}\cdot\bm{\sigma}_{2})(\bm{\tau}_{1}\cdot\bm{\tau}_{2})\\
 &  & +C_{5}\frac{i}{2}(\bm{q}\times\bm{k})\cdot(\bm{\sigma}_{1}+\bm{\sigma}_{2})+C_{6}(\bm{\sigma}_{1}\cdot\bm{q})(\bm{\sigma}_{2}\cdot\bm{q})\\
 &  & \left.+C_{7}(\bm{\sigma}_{1}\cdot\bm{q})(\bm{\sigma}_{2}\cdot\bm{q})(\bm{\tau}_{1}\cdot\bm{\tau}_{2})\right]f_{{\rm 2N}}(p_{1},p_{2},p_{1}^{\prime},p_{2}^{\prime}),
\end{eqnarray*}
where $\bm{\sigma}_{1,2}(\bm{\tau}_{1,2})$ are spin (isospin) matrices,
$B_{i},C_{i}$ are low-energy constants (LECs), $\bm{p}$ ($\bm{p}^{\prime}$)
is the relative incoming (outgoing) momentum, $\bm{q}=\bm{p}-\bm{p}^{\prime}$,
$\bm{k}=(\bm{p}+\bm{p}^{\prime})/2$ are momentum transfers, $\bm{p}_{i}$
and $\bm{p}_{i}^{\prime}$ are the momenta of the individual nucleons.
We adopt a single-particle regulator $f_{{\rm 2N}}=\prod_{i=1}^{2}g_{\Lambda}(p_{i})g_{\Lambda}(p_{i}^{\prime})$
where $g_{\Lambda}(p)=\exp(-p^{6}/2\Lambda^{6})$ is a soft cutoff
function. Similarly, for the three-nucleon force (3NF) we take a simple
contact term
\[
V_{{\rm 3N}}=C_{{\rm 3N}}f_{{\rm 3N}}(p_{1},\cdots,p_{3}^{\prime})
\]
where $C_{{\rm 3N}}$ is a LEC, $f_{{\rm 3N}}=\prod_{i=1}^{3}g_{\Lambda}(p_{i})g_{\Lambda}(p_{i}^{\prime})$
is a non-local single-particle regulator. The long-range pieces $V_{{\rm C}}$
and $V_{{\rm OPEP}}$ are both local potentials that do not depend
on $\Lambda$. In what follows we solve the nuclear ground state energies
using lattice EFT techniques\cite{Lee2009_PPNP-1,Elhatisari2016PRL,Lu2019PLB,Elhatisari2023arXiv}.
Particularly, we use Lanczos algorithm for $^{3}$H and the perturbative
quantum Monte Carlo method\cite{Lu2022PRL-1} for $^{4}$He and $^{16}$O.
The latter two are tightly bound doubly-magic nuclei for which the
perturbation theory is most accurate. See Supplemental Material\cite{SupMat}
for futher details of the interactions, methods and error quantifications.

\begin{figure}
\begin{centering}
\includegraphics[width=1\columnwidth]{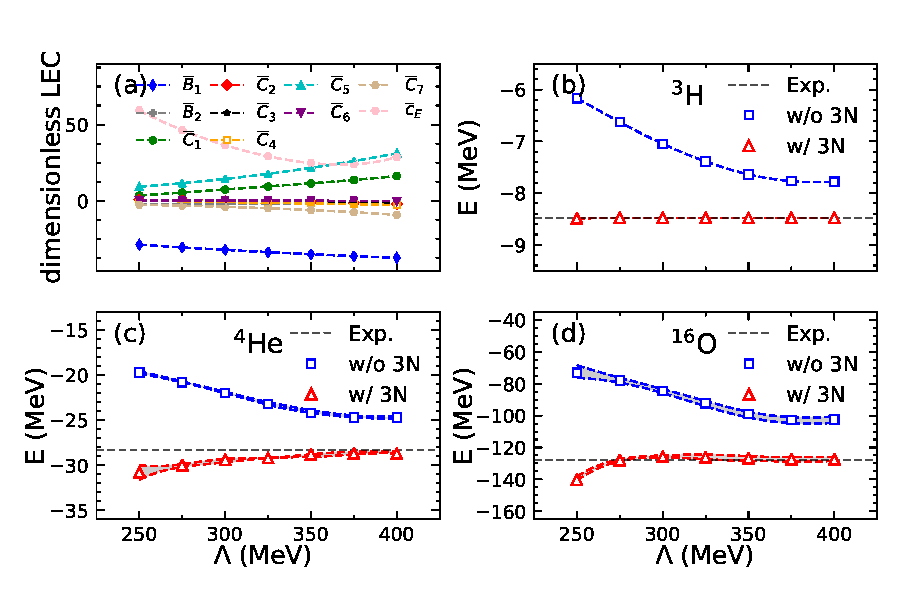}
\par\end{centering}
\caption{\label{fig:combined_figure_RGI}(a) Dimensionless LECs as functions
of $\Lambda$. (b) Triton binding energies calculated without (blue
squares) and with (red triangles) the three-body force $V_{{\rm 3N}}$.
The horizontal line denote the experimental value. (c) The same as
Fig.b but for $^{4}$He. The Monte Carlo statistical errors are smaller
than the symbols. The grey bands signify the estimated truncation
errors from the perturbative QMC calculations\cite{SupMat}. (d) The
same as Fig.c but for $^{16}$O.}
\end{figure}

Now we examine the Wilsonian RG flow by varying the cutoff $\Lambda$
from $400$ MeV to $250$ MeV with a constant interval of 25 MeV.
For each value of $\Lambda$ we determine the LECs $B_{i}$, $C_{i}$
and $C_{{\rm 3N}}$ by fitting to the experimental neutron-proton
scattering phase shifts for relative momenta below 200 MeV and the
triton binding energy $E(^{3}$H)=$-8.482$ MeV\cite{Alarcon2017EPJA,LiNing2018PRC}.
Note that the characteristic momentum in a typical nucleus can be
estimated as $Q\sim\sqrt{mE_{B}}\approx100$ MeV, where $E_{B}\approx7$
MeV is the single-nucleon binding energy. To confirm that fitting
with such low-momenta is enough, we have also fitted to the phase
shifts up to $250$ MeV and found no meaningful improvement for the
ground state calculations. According to the conventions in the RG
analysis, the running LECs can be combined with powers of $\Lambda$
to form dimensionless coupling constants $\overline{B}_{i}=m\Lambda B_{i}$,
$\overline{C}_{i}=m\Lambda^{3}C_{i}$ and $\overline{C}_{{\rm 3N}}=m\Lambda^{4}C_{{\rm 3N}}$.
In Fig.\ref{fig:combined_figure_RGI}(a) we show the dependence of
these parameters on $\Lambda$. Here we focus on the infrared asymptotic
behaviour and can observe three different tendencies. While $\overline{B}_{1}$
is approximately a constant and $\overline{C}_{{\rm 3N}}$ increases,
all other dimensionless LECs approach zero for $\Lambda\rightarrow0$.
In the language of the RG we can qualitatively say that $B_{1}$,
$C_{{\rm 3N}}$ and all other LECs correspond to marginal, relevant
and irrelevant operators, respectively. 

\begin{figure}
\begin{centering}
\includegraphics[width=0.85\columnwidth]{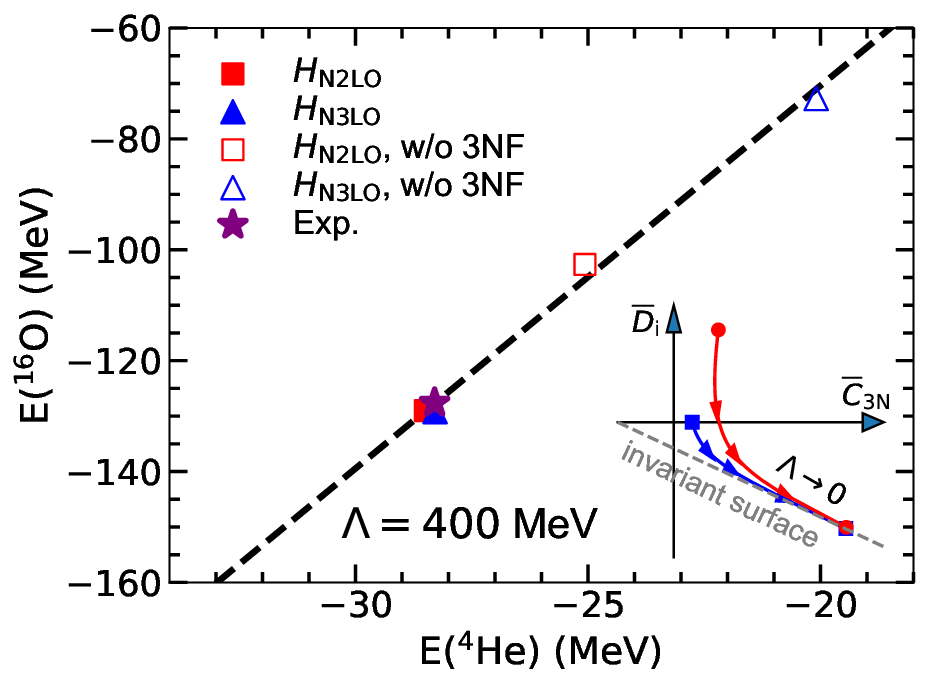}
\par\end{centering}
\caption{\label{fig:He4_O16_corr}The ground state energies of $^{4}$He and
$^{16}$O calculated with $H_{{\rm N2LO}}$ (squares) and $H_{{\rm N3LO}}$
(triangles). The results without the 3NF are denoted as open symbols.
The dashed line is for guiding the eyes. The star denotes the experimental
values. The inset schematizes two typical RG flows in the parameter
space. $\overline{C}_{{\rm 3N}}$ and $\overline{D}_{i}$ corresponds
to the 3NF and an irrelevant operator, respectively. As $\Lambda$
is lowered, two different theories with $\overline{D}_{i}=0$ and
$\overline{D}_{i}\protect\neq0$ flow to the same low-momentum EFT.}
\end{figure}

In Fig.\ref{fig:combined_figure_RGI}(b-d) we show the binding energies
of $^{3}$H, $^{4}$He and $^{16}$O calculated with running $\Lambda$
and the corresponding LECs. To see the effects of the 3NF we display
the results with and without the 3NF as triangles and squares, respectively.
The binding energies without the 3NF show appreciable $\Lambda$-dependence
and vary by about $20\%$ for the considered interval of $\Lambda$.
For the calculations with the 3NF, the $^{3}$H energy is fitted to
and always equals to the experimental value, whereas both the $^{4}$He
and $^{16}$O energies are predictions. Surprisingly, the simple 3NF
determined in $^{3}$H cancels most of the $\Lambda$-dependence in
$^{4}$He and $^{16}$O energies and significantly improves the description
of both experimental values. In particular, for $\Lambda\geq300$
MeV the $^{16}$O energy converges to the experimental value with
a discrepancy smaller than $2$ MeV. Note that our calculation has
no adjustable parameter other than the LECs fitted to two- and three-body
observables, thus the predictions are really model-independent.

The remaining $\Lambda$-dependence has multiple sources. For instance,
we have employed a soft cutoff function which does not completely
exclude the high-momentum subspace\cite{Bogner2006PLB,Bogner2007NPA}.
Another important source is the ommision of the contact terms that
explicitly violate the Galilean invariance\cite{LiNing2019PRC}. Generally,
these effects scale as negative powers of $\Lambda$, thus to extract
unambiguous predictions we need to make an extrapolation using multiple
values of $\Lambda$. This can be clearly seen for the convergence
pattern of $E(^{4}$He$)$ in Fig.\ref{fig:combined_figure_RGI}(c).
Following the Symanzik improvement program\cite{Symanzik1983NPB1,Symanzik1983NPB2}
in the lattice QCD, we can also add irrelavent operators to absorb
these regulator artifacts. Fortunately, in our cases such RG-invariance
breaking effects are sufficiently weak and we leave it as is.

Now we check whether the agreement with the experiments persists for
higher-order interactions. We fix $\Lambda$ to 400 MeV that minimizes
the regulator artifacts, add fifteen two-body contact terms at $\mathcal{O}(Q^{4})$
to the Hamiltonian and refit all LECs using the same prescription
as for $H_{{\rm N2LO}}$. We denote the resulting Hamiltonian as $H_{{\rm N3LO}}$\cite{SupMat}.
Note that $H_{{\rm N2LO}}$ and $H_{{\rm N3LO}}$ give nearly the
same phase shifts below 200 MeV and the same triton binding energy.
In Fig.\ref{fig:He4_O16_corr} we plot $E$($^{4}$He) versus $E$($^{16}$O)
predicted by both Hamiltonians. We also display the results without
the 3NF as open symbols. The star denotes the experimental values.
We see that the $^{4}$He and $^{16}$O energies are highly correlated
that their energy corrections from the higher-order 2NF can be canceled
simultaneously by adjusting the 3NF. We also found that the $^{3}$H
and $^{4}$He energies are correlated as expected by the Tjon line\cite{Tjon1975PLB}.
Thus we conclude that at N$^{3}$LO a single 3NF is also enough for
reproducing all three binding energies.

The phenomena that a simple pattern emerges from complex underlying
interactions is known as the universality. Famous examples in nuclear
physics are the Tjon line\cite{Tjon1975PLB} and Coester line\cite{Coester1970PRC}.
Similar to the cases in thermodynamics, such emergence behaviour in
nuclear physics can also be understood via the RG transformations.
In the inset of Fig.\ref{fig:He4_O16_corr} we schematize the Wilsonian
RG flows of two Hamiltonians starting with and without a typical 2NF
at $\mathcal{O}(Q^{4})$. The 2NF is represented by a single parameter
$\overline{D}_{i}$. Note that $\overline{C}_{{\rm 3N}}$ and $\overline{D}_{i}$
correspond to relevant and irrelevant operators, respectively. Following
the Wilsonian RG analysis near a fixed point, for any value of $\overline{D}_{i}$
we can always adjust $\overline{C}_{{\rm 3N}}$ to make the RG flow
converge to the same low-momentum EFT located on a low-dimensional
invariant surface. These RG flows are independent of the nucleon number
and induce the observed correlations between the binding energies.

\paragraph{Summary and perspective}

We present a general framework for buiding RG invariant EFTs. In brief,
our method is a non-relativistic counterpart of the Wilsonian RG theory.
The only requirement is to regulate all single-particle momenta with
the same cutoff. It applies for non-perturbative as well as perturbative
EFTs. The RG flows can be calculated explicitly from the operator
commutators. The first application of this paradigm to realistic chiral
forces yields promising results. Our findings suggest that the RG
invariance might be realized at much lower orders and smaller cutoffs
than previously expected. Its consequences on other $\chi$EFT calculations
are being examined. As the rigidity of the predictions is guaranteed
by the RG invariance, fine-tuning is not allowed and we expect that
the discrepancies to the experiments would signify the real missing
physics such as the medium effects.
\begin{acknowledgments}
We are grateful for discussions with S.-G. Zhou and members of the
Nuclear Lattice Effective Field Theory Collaboration. We gratefully
acknowledge funding by NSAF (Grant No. U1930403) and the National
Natural Science Foundation of China (Grant Nos. 12275259, 11961141004,
12070131001, 12047503) as well as computational resources provided
by the Beijing Super Cloud Computing Center (BSCC) and TianHe 3F.
\end{acknowledgments}

\begin{widetext}

\newpage



\part*{Supplemental Material}

In the main text we present a numerical demonstration using the realistic
chiral interactions. Here we present the details of the chiral interaction,
the many-body methods and the error quantifications. Note that some
of the materials have already been given in Ref.\cite{Lu2019_PLB}
and the supplemental materials of Ref.\cite{Lu2022_PRL}. For completeness
we also include them here. We also give a proof of the convergence
of the flow equation.

\subsection{Nuclear chiral interactions}

Up to N$^{2}$LO the form of the interaction is exactly the same as
that we used in Ref.\cite{Lu2022_PRL}. For contact terms we employ
the semi-local operator basis that contains isospin dependent terms
proportional to $\bm{\tau}_{1}\cdot\bm{\tau}_{2}$. In this form,
the contact terms are mostly seperable and can be written as products
of one-body density operators, which is amenable to quantum Monte
Carlo simulations. We note that similar constructions have already
been used for the N$^{2}$LO Green's function Monte Carlo calculations\cite{Gezerlis2013,Lonardoni2018}.
In addition, we also present the fifteen semi-local N$^{3}$LO contact
terms.

In momentum space the contact terms up to $\mathcal{O}(Q^{4})$ are
written as 
\begin{eqnarray}
V_{Q^{0}} & = & B_{1}+B_{2}(\bm{\sigma}_{1}\cdot\bm{\sigma}_{2})\nonumber \\
V_{{\rm Q^{2}}} & = & C_{1}q^{2}+C_{2}q^{2}(\bm{\tau}_{1}\cdot\bm{\tau}_{2})+C_{3}q^{2}(\bm{\sigma}_{1}\cdot\bm{\sigma}_{2})+C_{4}q^{2}(\bm{\sigma}_{1}\cdot\bm{\sigma}_{2})(\bm{\tau}_{1}\cdot\bm{\tau}_{2})\nonumber \\
 &  & +C_{5}\frac{i}{2}(\bm{q}\times\bm{k})\cdot(\bm{\sigma}_{1}+\bm{\sigma}_{2})+C_{6}(\bm{\sigma}_{1}\cdot\bm{q})(\bm{\sigma}_{2}\cdot\bm{q})+C_{7}(\bm{\sigma}_{1}\cdot\bm{q})(\bm{\sigma}_{2}\cdot\bm{q})(\bm{\tau}_{1}\cdot\bm{\tau}_{2})\nonumber \\
V_{{\rm Q^{4}}} & = & D_{1}q^{4}+D_{2}q^{4}(\bm{\tau}_{1}\cdot\bm{\tau}_{2})+D_{3}q^{4}(\bm{\sigma}_{1}\cdot\bm{\sigma}_{2})+D_{4}q^{4}(\bm{\sigma}_{1}\cdot\bm{\sigma}_{2})(\bm{\tau}_{1}\cdot\bm{\tau}_{2})\nonumber \\
 &  & +D_{5}q^{2}(\bm{\sigma}_{1}\cdot\bm{q})(\bm{\sigma}_{2}\cdot\bm{q})+D_{6}q^{2}(\bm{\sigma}_{1}\cdot\bm{q})(\bm{\sigma}_{2}\cdot\bm{q})(\bm{\tau}_{1}\cdot\bm{\tau}_{2})+D_{7}q^{2}k^{2}\nonumber \\
 &  & +D_{8}q^{2}k^{2}(\bm{\sigma}_{1}\cdot\bm{\sigma}_{2})+D_{9}(\bm{q}\cdot\bm{k})^{2}+D_{10}(\bm{q}\cdot\bm{k})^{2}(\bm{\sigma}_{1}\cdot\bm{\sigma}_{2})\nonumber \\
 &  & +D_{11}\frac{i}{2}q^{2}(\bm{q}\times\bm{k})\cdot(\bm{\sigma}_{1}+\bm{\sigma}_{2})+D_{12}\frac{i}{2}q^{2}(\bm{q}\times\bm{k})\cdot(\bm{\sigma}_{1}+\bm{\sigma}_{2})(\bm{\tau}_{1}\cdot\bm{\tau}_{2})\nonumber \\
 &  & +D_{13}k^{2}(\bm{\sigma}_{1}\cdot\bm{q})(\bm{\sigma}_{2}\cdot\bm{q})+D_{14}q^{2}(\bm{\sigma}_{1}\cdot\bm{k})(\bm{\sigma}_{2}\cdot\bm{k})\nonumber \\
 &  & +D_{15}\left[(\bm{q}\times\bm{k})\cdot\bm{\sigma}_{1}\right]\left[(\bm{q}\times\bm{k})\cdot\bm{\sigma}_{2}\right],\label{eq:contact_terms_Q4}
\end{eqnarray}
where $B_{1-2}$, $C_{1-7}$ and $D_{1-15}$ are low-energy constants
(LECs), $\bm{q}=\bm{p}^{\prime}-\bm{p}$ and $\bm{k}=(\bm{p}+\bm{p}^{\prime})/2$
are momentum transfers, $\bm{\sigma}_{1,2}$ and $\bm{\tau}_{1,2}$
are spin and isospin Pauli matrices, respectively. Due to wave function
anti-symmetrization this choice of operators is not unique. For convenience
in lattice calculations we choose the semi-local basis in Eq.(\ref{eq:contact_terms_Q4})
that minimizes the dependence on $\bm{k}$. In Ref.\cite{Reinert2018}
it was pointed out that at order $Q^{4}$ there are two redundant
operators in reproducing the N-N phase shifts. This correspond to
the fact that the operators attached to $D_{9}$ and $D_{10}$ vanish
for on-shell momenta $|\bm{p}|=|\bm{p}^{\prime}|$. 

For all two-body contact terms we introduce a non-local regulator
$f_{{\rm 2N}}=\exp[-\sum_{i=1}^{2}\left(p_{i}^{6}+p_{i}^{\prime6}\right)/(2\Lambda^{6})]$,
where $\bm{p}_{1,2}$ and $\bm{p}_{1,2}^{\prime}$ are incoming and
outgoing single particle momenta, respectively. Comparing with the
form used in Ref.\cite{Lu2022_PRL}, here we insert an extra factor
$2$ before $\Lambda$ for convenience. For two colliding nuclei in
the center of mass frame, we have the relative momenta $\bm{p=\bm{p}_{1}=-\bm{p}_{2}}$
and $\bm{p}^{\prime}=\bm{p}_{1}^{\prime}=-\bm{p}_{2}^{\prime}$, thus
$f_{{\rm 2N}}=\exp(-(p^{6}+p^{\prime6})/\Lambda^{6})$ reduces to
the non-local regulator extensively used in chiral EFT constructions\cite{Epelbaum2009,Machleidt2011}.
However, for many-body systems, our regulator is imposed on single-particle
momenta rather than relative momenta, thus gives different predictions
even though the two-body LECs are the same.

In the main text we include a long-range one-pion-exchange potential
(OPEP) $V_{{\rm OPEP}}$ in the chiral interactions. Here we give
the details. In momentum space we have
\[
V_{{\rm OPEP}}=-\frac{g_{A}^{2}f_{\pi}(q^{2})}{4F_{\pi}^{2}}\left[\frac{(\bm{\sigma}_{1}\cdot\bm{q})(\bm{\sigma}_{2}\cdot\bm{q})}{q^{2}+M_{\pi}^{2}}+C_{\pi}^{\prime}\bm{\sigma}_{1}\cdot\bm{\sigma}_{2}\right](\bm{\tau}_{1}\cdot\bm{\tau}_{2}),
\]
where $g_{A}=1.287$, $F_{\pi}=92.2$ MeV, $M_{\pi}=134.98$ MeV are
the axial-vector coupling constant, pion decay constant and pion mass,
respectively. The OPEP is regulated with a local exponential regulator
$f_{\pi}(q^{2})=\exp[-(q^{2}+M_{\pi}^{2})/\Lambda_{\pi}^{2}]$ with
$\Lambda_{\pi}$ fixed to $300$MeV. We also tested other values of
$\Lambda_{\pi}$ and found similar results as presented here. The
constant $C_{\pi}^{\prime}$ is defined as 
\begin{align*}
C_{\pi}^{\prime} & =\frac{1}{3\Lambda_{\pi}^{3}}\Biggl[\Lambda_{\pi}(\Lambda_{\pi}^{2}-2M_{\pi}^{2})+2\sqrt{\pi}M_{\pi}^{3}\exp(\frac{M_{\pi}^{2}}{\Lambda_{\pi}^{2}}){\rm erfc}(\frac{M_{\pi}}{\Lambda_{\pi}})\Biggr].
\end{align*}
 The term proportional to $C_{\pi}^{\prime}$ is a counterterm introduced
to remove the short-range singularity from the OPEP\cite{Reinert2018}.
We note that the OPEP regulated in this way is soft and adaptive to
perturbative calculations\cite{Lu2022_PRL}.

For the three-body force at N$^{2}$LO we adopt a simple 3N contact
term with Wigner SU(4) symmetry\cite{Wigner1937}, 
\[
V_{{\rm 3N}}=C_{{\rm 3N}}f_{{\rm 3N}}(p_{1},p_{2},\cdots,p_{3}^{\prime}),
\]
where $C_{{\rm 3N}}$ is a LEC, $f_{{\rm 3N}}=\exp[-\sum_{i=1}^{3}\left(p_{i}^{6}+p_{i}^{\prime6}\right)/(2\Lambda^{6})]$
is a seperable single-particle regulator. Note that we use the same
soft cutoff functions and the same value of $\Lambda$ in $f_{{\rm 2N}}$
and $f_{{\rm 3N}}$.

Besides the nuclear force we also include a static Coulomb force,
\[
V_{{\rm C}}=:\frac{\alpha}{2}\int d^{3}\bm{r}_{1}d^{3}\bm{r}_{2}\frac{\rho_{p}(\bm{r}_{1})\rho_{p}(\bm{r}_{2})}{|\bm{r}_{1}-\bm{r}_{2}|}:
\]
 where $\rho_{p}$ is the total proton density, $\alpha=1/137$ is
the fine structure constant. 

We implement the Hamiltonian on a cubic lattice using fast Fourier
transform and determine the LECs $B_{i}$, $C_{i}$, $D_{i}$ and
$C_{{\rm 3N}}$ by fitting to the low-energy NN phase shifts, mixing
angles and triton energy. The method is based on Ref.\cite{Lu2016}.
We decompose the scattering waves on the lattice into different partial
waves, then employ the real and complex auxiliary potentials to extract
the asymptotic radial wave functions. We follow the conventional procedure
for fitting the LECs in the continuum\cite{Epelbaum2005}. We first
determine the spectroscopic LECs for each partial wave, then the $B_{i}$,
$C_{i}$ and $D_{i}$ can be obtained by solving the linear equations. 

\begin{figure}
\begin{centering}
\includegraphics[width=1\textwidth]{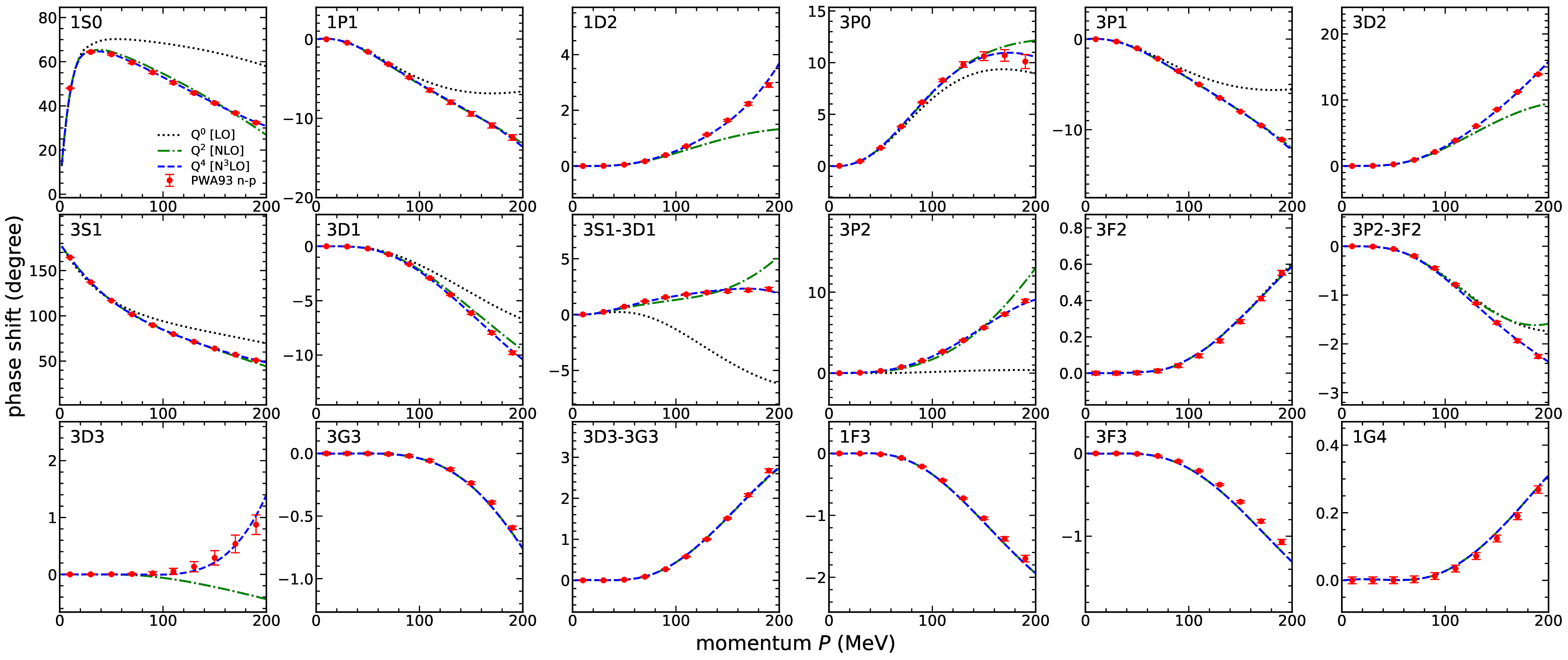}
\par\end{centering}
\caption{\label{fig:Calculated-phase-shifts}Neutron-proton phase shifts calculated
with the momentum cutoff $\Lambda$=400 MeV. Dotted, dash-dotted and
dashed lines denote the LO, NLO and N$^{3}$LO results, respectively.
Red circles with error bars show the empirical values\cite{Stoks1993}.}

\end{figure}

In this work we consider NN scattering up to a relative momentum $P_{{\rm rel}}=200$
MeV. In Fig.\ref{fig:Calculated-phase-shifts} we show the calculated
phase shifts for $\Lambda=400$ MeV. The dotted, dash-dotted and dashed
lines denote the results at LO, NLO and N$^{3}$LO, respectively.
The red dots with error bars are empirical values from the Nijmegen
partial wave analysis (NPWA)\cite{Stoks1993}.

In principle the $\pi$-$N$ coupling $g_{A}$ should also run with
$\Lambda$. However, as the OPEP does not explicitly contain $\Lambda$,
we expect that the dependence is rather weak. We confirm this point
by refitting the LECs with different values of $g_{A}$. We found
that $g_{A}=1.287$ always approximately gives the best fit for $\Lambda=250\simeq400$
MeV and conclude that the OPEP do not evolve.

Compared with other high-precision N$^{3}$LO chiral interactions,
our construction lacks the two-pion exchange potentials (TPEP). The
reason is that for momentum much lower than $2M_{\pi}\approx270$
MeV, the TPEP can be approximately absorbed into the contact terms
and discarded. This point can be clearly seen by inspecting the peripheral
partial waves like $^{3}F_{3}$ or $^{1}G_{4}$ in Fig.\ref{fig:Calculated-phase-shifts}.
These partial waves are solely determined by the OPEP. The good reproduction
of the empirical values makes it unnecessary to include the TPEP in
this momentum interval. This argument also applies to three-pion exchange
potentials.

In the three-body sector we also ommitted the one-pion-exchange three-body
force. It was long known that the two three-body terms at N$^{2}$LO
can not be uniquely fixed with $^{3}$H and $^{4}$He binding energies
which are highly correlated\cite{Platter2005,Klein2018}. In this
work we also found similar correlation between the $^{4}$He and $^{16}$O
energies for the three-body forces. The situation is quite similar
to the case of N$^{3}$LO contact terms. Thus we still can not distinguish
the two three-body forces even with the $^{16}$O binding energy as
an additional input. This finding can be contrasted with the recent
practices of adapting the three-body forces to many-body observables.
We conjecture that the difference is related to the different regulators
and need further investigations.

Finally, the RG transform induces additional terms such as $Q^{2}$,
$Q^{2}(\bm{\sigma}_{1}\cdot\bm{\sigma}_{2})$, $(\bm{Q}\cdot\bm{\sigma}_{1})(\bm{Q}\cdot\bm{\sigma}_{2})$
etc., where $\bm{Q}=\bm{p}_{1}+\bm{p}_{2}=\bm{p}_{1}^{\prime}+\bm{p}_{2}^{\prime}$
is the center of mass momentum. For a full RG invariance calculation
we need to include all of them in the expansion of the EFT Hamiltonian,
then fit the corresponding LECs to restore the Galillean invariance
in each partial waves. This is as difficult as fitting the normal
LECs and only has been performed for the leading order terms\cite{LiNing2019}.
Fortunately, since the Galillean invariance breaking effects originate
from the single-particle regulator imposed at a large momentum $\Lambda$,
the LECs for these Galillean-invariance-restoration (GIR) terms scale
as negative powers of $\Lambda$ and decay to zero for large $\Lambda$.
For example, the calculations in the main text simply ommit all GIR
terms and we can expect that the results converge as $\mathcal{O}(\Lambda^{-2})$
for large $\Lambda$. By including the GIR terms at $\mathcal{O}(Q^{2})$
we can accelerate the convergence speed to $\mathcal{O}(\Lambda^{-4})$.
This procedure is a non-relativistic counterpart of the Symanzik improvement
for the lattice QCD action\cite{Symanzik1983a,Symanzik1983b}, where
irrelevant operators proportional to $a^{2}$, $a^{4}$, etc., are
added to the QCD Lagrangian to absorb the lattice artifacts, where
$a$ is the lattice spacing. The most notable difference is that in
an EFT we can not take the limit $\Lambda\rightarrow\infty$ as new
physics may emerge at certain hard scale where the EFT breaks down. 

\subsection*{Perturbative quantum Monte Carlo method}

In this work we solve the $^{4}$He and $^{16}$O ground state energies
using the perturbative quantum Monte Carlo method \cite{Lu2022_PRL}.
The starting point is a zeroth order Hamiltonian respecting the Wigner
SU(4) symmetry\cite{Wigner1937}. While the SU(4) Hamiltonian in Ref.\cite{Lu2022_PRL}
is designed for lattice spacing $a=1.32$ fm, in this work we use
a finer lattice with $a=1.0$ fm and slightly adjust the form of the
interaction. On a pariodic $L^{3}$ cube with lattice coordinates
$\bm{n}=(n_{x},n_{y},n_{z})$, the zeroth order Hamiltonian is
\begin{equation}
H_{0}=K+\frac{1}{2}C_{{\rm SU4}}\sum_{\bm{n}}:\tilde{\rho}^{2}(\bm{n}):,\label{eq:SU(4)Hamiltonian}
\end{equation}
where $K$ is the kinetic energy term with nucleon mass $m=938.92$
MeV and the $::$ symbol indicate normal ordering. The smeared density
operator $\tilde{\rho}(\bm{n})$ is defined as
\[
\tilde{\rho}(\bm{n})=\sum_{i}\tilde{a}_{i}^{\dagger}(\bm{n})\tilde{a}_{i}(\bm{n})+s_{L}\sum_{|\bm{n}^{\prime}-\bm{n}|=1}\sum_{i}\tilde{a}_{i}^{\dagger}(\bm{n}^{\prime})\tilde{a}_{i}(\bm{n}^{\prime}),
\]
where $i$ is the joint spin-isospin index and the smeared annihilation
and creation operators are defined as
\[
\tilde{a}_{i}(\bm{n})=\sum_{\bm{n}^{\prime}}g(|\bm{n}-\bm{n}^{\prime}|)a_{i}(\bm{n}^{\prime}),
\]
where $g(r)=(\Lambda_{NL}/\sqrt{2\pi})^{3}\exp(-\Lambda_{NL}^{2}r^{2}/2)$
is a Gaussian smearing function. The summation over the spin and isospin
implies that the interaction is SU(4) invariant. The parameter $s_{L}$
controls the strength of the local part of the interaction, while
$\Lambda_{NL}$ controls the strength of the non-local part of the
interaction. Here we include both kinds of smearing. Both $s_{L}$
and $\Lambda_{NL}$ have an impact on the range of the interactions.
The parameter $C_{{\rm SU4}}$ gives the strength of the two-body
interactions. In this work we use the parameter set $C_{{\rm SU4}}=-3.7912\times10^{-5}$
MeV$^{-2}$, $s_{L}=0.182$ and $\Lambda_{NL}=300$ MeV. These parameters
together with a properly chosen three-body force reproduce the binding
energies of light nuclei.

In auxiliary field Monte Carlo calculations, the SU(4) Hamiltonian
in Eq.(\ref{eq:SU(4)Hamiltonian}) generates the same auxiliary fields
for spin-up and spin-down particles. For even-even nuclei like $^{4}$He
and $^{16}$O, the determinant from the wave function antisymmmetrization
can be factorized into a product of identical determinants, which
is positive definite. In such a sign-problem-free scenario the Monte
Carlo simulation can be very accurate\cite{Lu2019_PLB}. However,
the realistic interactions usually break the SU(4) symmetry and induce
severe sign problem, causing large statistical errors. To solve this
problem, we rewrite the full Hamiltonian $H$ as a sum of the SU(4)
symmetric Hamiltonian $H_{0}$ and a residual term $\delta H$, then
use Monte Carlo techinques to simulate $H_{0}$ non-perturbatively
and incoorporate $\delta H$ using the 2nd order perturbation theory.
In this way we can largely suppress the sign problem at the price
of introducing minor perturbative truncation errors\cite{Lu2022_PRL}.
In this work $H$ can be either the realistic chiral Hamiltonian $H_{{\rm N2LO}}$
fitted with $\Lambda=250\simeq400$ MeV or $H_{{\rm N3LO}}$ fitted
with $\Lambda=400$ MeV. 

\subsection*{Error quantifications}

For a solid demonstration, we need to estimate the numerical errors
due to the many-body methods. Note that we only consider the errors
from solving a given Hamiltonian. The uncertainties from truncating
the chiral Hamiltonian at a given order are reflected by the breaking
of the RG invariance, which has been discussed in the main text.

\begin{figure}
\begin{centering}
\includegraphics[width=1\textwidth]{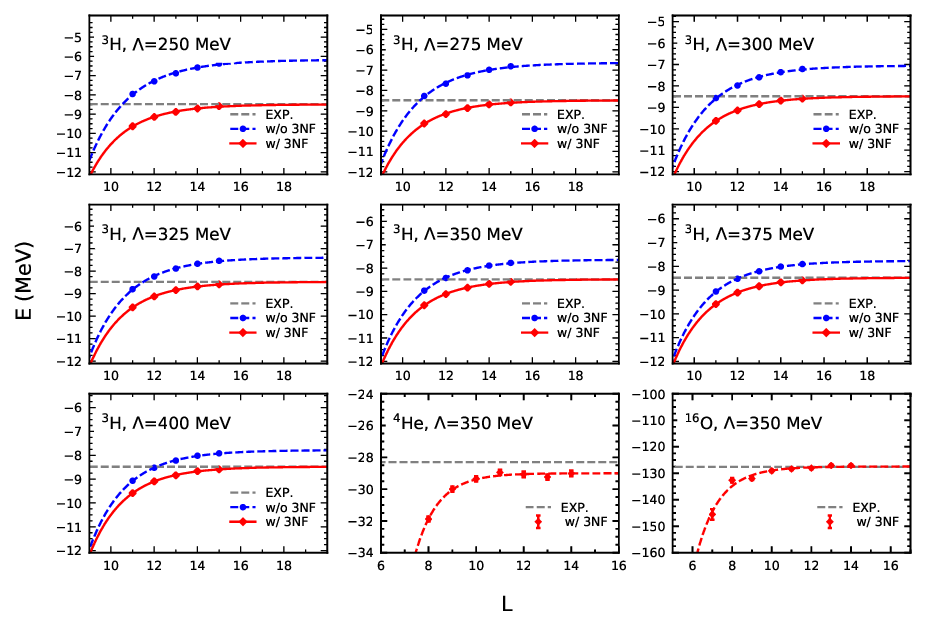}
\par\end{centering}
\caption{\label{fig:InfiniteVolume}The ground state energies of $^{3}$H,
$^{4}$He and $^{16}$O calculated with different box size $L$. The
circles and diamonds denote the results with and without the three-body
force, respectively. The dashed and solid lines are fitted curves
for extrapolating to the infinite volume limit. The horizontal lines
are corresponding experimental values. The errorbars for $^{4}$He
and $^{16}$O denote the Monte Carlo statistical errors. }
\end{figure}

\paragraph*{Continuum limit}

In this work we fix the lattice spacing to $a=1.0$ fm. This corresponds
to a momentum cutoff $\Lambda_{a}=\pi/a\approx620$ MeV. When we consider
a Hamiltonian regulated by an extra regulator with a cutoff $\Lambda=250\simeq400$
MeV $\ll\Lambda_{a}$, the spatial lattice is invisible to the particles.
In other words, we can further take the continuum limit $a\rightarrow0$
or $\Lambda_{a}\rightarrow\infty$ to eliminate the discretization
errors. In this sense the lattice is used as a numerical tool for
solving the quantum many-body problem and has no physical meaning.
What is physical relevance is the cutoff $\Lambda$ defined in the
spatial continuum. Thus our results can be immediately benchmarked
using other many-body algorithms designed for general Hamiltonians,
e.g., no-core shell model\cite{Barrett2013} or IM-SRG method\cite{Hergert2016}.
This setting is quite different from the recent lattice EFT calculations,
where the lattice was explicitly used as the momentum regulator and
the lattice spacing must be kept finite throughout.

\paragraph*{Infinite volume limit}

To make a precision fit for the three-body parameter $C_{{\rm 3N}}$,
we calculate the triton binding energy by directly diagonalizing the
lattice Hamiltonian using the Lanczos method. As this method becomes
expensive for large volumes, we use small box sizes $L=11\simeq15$
and extrapolate the results to $L\rightarrow\infty$. We measure the
box size in unit of the lattice spacing. In Fig.\ref{fig:InfiniteVolume}
we show the triton binding energies calculated with and without the
three-body force as functions of $L$. The dashed and solid lines
are fitted curves using the ansatz for three particles in the unitary
limit\cite{Meissner2015_PRL},
\[
E(L)=E(\infty)+\mathcal{A}\exp(2\kappa L/\sqrt{3})/(\kappa L)^{3/2}
\]
with $\kappa=\sqrt{-mE(\infty)}$, $m$ is the nucleon mass. We ensure
that all triton energies with the three-body force converge to the
experimental value $E(^{3}$H)=$-8.482$ MeV. In Fig.2b of the main
text, the displayed triton energies are extrapolated values.

For $^{4}$He we use the Monte Carlo method and take $L=12$ for all
calculations. For $^{16}$O we take $L=11$. For saving the computational
resources, we only examine the convergence patterns for $\Lambda=350$
MeV. The results are also displayed in Fig.\ref{fig:InfiniteVolume}.
Clearly the results with $L=12$ and $11$ are sufficiently close
to the infinite volume limit.

\paragraph*{Perturbative truncation errors}

For $^{4}$He and $^{16}$O energies we employed the 2nd order perturbation
theory\cite{Lu2022_PRL}. The errors from truncating the perturbative
series account for most of the numerical errors. Following Ref.\cite{Lu2022_PRL},
we roughly estimate the truncation errors by varying the zeroth order
Hamiltonian $H_{0}$. If the perturbative series converge and we can
calculate up to infinite orders, the results will be determined by
$H$ and independent of $H_{0}$. Thus any dependence on $H_{0}$
is a signature of truncating the series at a finite order. For all
Monte Carlo calculations in the main text, we perform the simulation
using three different $H_{0}$ with $C_{{\rm SU4}}$ multiplied by
$0.9$, $1.0$ and $1.1$, respectively. We found that the eigenvalues
of $H_{0}$ vary by about 20\% for different values of $C_{{\rm SU4}}$.
However, after including $\delta H=H-H_{0}$ using the perturbative
QMC method, we found that the energies are almost constants against
$C_{{\rm SU4}}$. In Fig.2 of the main text we plot the uncertainties
due to the variation of $C_{{\rm SU4}}$ with the grey bands. 

\subsection*{Convergence of the flow equation}

Below the flow Eq.(4) in the main text we claim that it converges
to the fixed point $H_{1}=\eta=0$ as long as $H_{1}$ is small. Here
we give a more detailed derivation. The proof follows the derivation
of the $V_{lowk}$ potential\cite{Bogner2010} with minor modifications.

In the main text we write the Hamiltonian as $H=H_{0}+H_{1}$ and
show that $H_{0}$ is block-diagonal. Formally we can write
\[
H_{0}=\left(\begin{array}{cc}
A_{0}\\
 & A_{1}
\end{array}\right),\qquad H_{1}=\left(\begin{array}{cc}
\epsilon & b\\
b^{\dagger}
\end{array}\right),
\]
where $A_{0}$, $A_{1}$ and $\epsilon$ are Hermite matrices. Then
we can calculate the generator as
\[
i\eta=[H_{0},H_{1}]=\left(\begin{array}{cc}
A_{0}\epsilon-\epsilon A_{0} & A_{0}b-bA_{1}\\
A_{1}b^{\dagger}-b^{\dagger}A_{0}
\end{array}\right),
\]
then substituting $i\eta$ into the flow equation we get 
\[
\frac{dH}{dt}=[i\eta,H]=[\left(\begin{array}{cc}
A_{0}\epsilon-\epsilon A_{0} & A_{0}b-bA_{1}\\
A_{1}b^{\dagger}-b^{\dagger}A_{0}
\end{array}\right),\left(\begin{array}{cc}
A_{0}+\epsilon & b\\
b^{\dagger} & A_{1}
\end{array}\right)].
\]
Taking the $(1,2)$ matrix element we get
\[
\frac{db}{dt}=(A_{0}\epsilon-\epsilon A_{0})b+(A_{0}b-bA_{1})A_{1}-(A_{0}+\epsilon)(A_{0}b-bA_{1})\approx2A_{0}bA_{1}-bA_{1}^{2}-A_{0}^{2}b,
\]
where we only keep terms linear in $b$ or $\epsilon$. These terms
are from $H_{1}$ which contains at least one $\delta\Psi$ or $\delta\Psi^{\dagger}$.
When $\Lambda^{\prime}$ is close to $\Lambda$ these terms are much
smaller than $\Psi$ and $\Psi^{\dagger}$, thus can be treated as
infinitesimals compared with $A_{0}$ and $A_{1}$. Then we can calculate
the derivative of the positive definite expression,
\begin{align*}
\frac{d}{dt}{\rm Tr}(b^{\dagger}b) & ={\rm Tr}(\frac{db^{\dagger}}{dt}b+b^{\dagger}\frac{db}{dt})=2{\rm Tr}(2A_{1}b^{\dagger}A_{0}b-A_{1}^{2}b^{\dagger}b-bb^{\dagger}A_{0}^{2})\\
 & =-2{\rm Tr}[(A_{0}b-bA_{1})^{\dagger}(A_{0}b-bA_{1})]\leq0,
\end{align*}
thus the off-diagonal blocks always decreases in the course of the
evolution. As every term in $H_{1}$ has non-zero matrix elements
in the off-diagonal block $b$ or $b^{\dagger}$, eliminating $b$
and $b^{\dagger}$also means getting rid of $H_{1}$.

\subsection*{}

\end{widetext}


\end{document}